\documentclass[a4paper,fleqn]{cas-sc}
\usepackage{soul}
\usepackage[sort&compress,numbers]{natbib}
\usepackage{lineno}
\usepackage{setspace}
\usepackage{graphics}
\usepackage{graphicx}
\usepackage{cuted}

\def\tsc#1{\csdef{#1}{\textsc{\lowercase{#1}}\xspace}}
\tsc{WGM}
\tsc{QE}
\tsc{EP}
\tsc{PMS}
\tsc{BEC}
\tsc{DE}

\begin{document}

\let\WriteBookmarks\relax
\def\floatpagepagefraction{1}
\def\textpagefraction{.001}
\shorttitle{On the Elastic Properties of Single-Walled Phagraphene Nanotubes}
\shortauthors{Pereira J\'unior \textit{et~al}.}

\title [mode = title]{On the Elastic Properties of Single-Walled Phagraphene Nanotubes}

\author[1]{M. L. Pereira J\'unior}
\author[2]{J. M. De Sousa}
\author[3]{W. H. S. Brand\~ao}
\author[3]{A. L. Aguiar}
\author[4]{R. A. Bizao}
\author[1,5]{L. A. Ribeiro J\'unior}
\cormark[1]
\ead{ribeirojr@unb.br}
\author[6,7]{D. S. Galv\~ao}

\address[1]{Institute of Physics, University of Bras\'ilia, 70910-900, Bras\'ilia, Brazil}
\address[2]{Federal Institute of Education, Science and Technology of Piau\'i, S\~ao Raimundo Nonato, Piau\'i, Brazil}
\address[3]{Physics Department, Piau\'i Federal University, Teresina, Piauí, 64049-550, Brazil}
\address[4]{Institute of Mathematics and Computer Sciences, University of S\~ao Paulo, S\~ao Carlos, S\~ao Paulo, Brazil.}
\address[5]{PPGCIMA, Campus Planaltina, University of Bras\'{i}lia, 73345-010, Bras\'{i}lia, Brazil}
\address[6]{Applied Physics Department, 'Gleb Wataghin' Institute of Physics, State University of Campinas, Campinas,SP, 13083-970, Brazil}
\address[7]{Center for Computing in Engineering \& Sciences, State University of Campinas, Campinas, SP, 13083-970, Brazil}

\cortext[cor1]{Corresponding author}

\begin{abstract}
Phagraphene (PhaG) is a quasi-planar 2D structure composed of $5-6-7$ ring sequence. We have investigated the structural and mechanical properties of phagraphene nanotubes (PhaNTs) through fully atomistic reactive molecular dynamics (MD) simulations. For comparison purposes, the results were also contrasted to similar carbon nanotubes (CNTs). Results showed that PhaNTs and CNTs present similar brittle fracture mechanisms. The Young's modulus values obtained for PhaNTs were smaller than the corresponding ones for CNTs. Both, PhaNTs and CNTs, present equivalent fracture strains ranging between 15\%-20\%. For the ultimate strength values, CNTs present values about 30\% higher than the corresponding ones for PhaNTs.       
\end{abstract}




\maketitle
\doublespacing

\section{Introduction}
\label{sec1}
The carbon ability to form different types of chemical bonds, with itself or other elements, has contributed to the advancement of organic-based nanomaterials in the last three decades \cite{frackowiak2001carbon,schlapbach2001hydrogen,zhang2009carbon,krasheninnikov2007engineering}. The discovery of fullerenes \cite{kroto} and carbon nanotubes (CNTs) \cite{iijima}, and more recently graphene \cite{novoselov2004electric}, illustrates the carbon importance \cite{allen2010honeycomb,sanchez2012biological}. In particular, the advent of graphene created a new era in materials science \cite{shao2010graphene,avouris2012graphene,kuila2012chemical,stoller2008graphene}. The graphene unique physical and chemical properties have been exploited in a variety of applications, such as biosensors \cite{shao2010graphene}, ultra-capacitors \cite{stoller2008graphene}, and photonics \cite{avouris2012graphene}. However, graphene is a zero-bandgap material which precludes its use in some electronics applications \cite{withers2010electron}. The search for structures that share some of remarkable graphene properties but with non-zero bandgap has renewed and/or created the interest in other 2D carbon allotropes.  Among the plethora of these structures, we can mention graphynes \cite{baughman1987structure}, pentagraphene \cite{zhang2015penta}, R-graphyne \cite{yin2013r}, porous graphene \cite{brunetto2012nonzero}, graphenylene \cite{yu2013graphenylene}, twin-graphene \cite{jiang2017twin}, $\Psi$-graphene \cite{psigraphene2017}, and phagraphene (PhaG) \cite{wang2015phagraphene,CITARNOSSOMRS2018}. PhaG is a quasi-planar 2D structure formed of sp$^{2}$-like hybridized carbon atoms with a $5-6-7$ ring sequence and its binding energy ($-9.03$ eV/atom) is close to that of graphene ($ -9.23$ eV/atom) \cite{wang2015phagraphene}.

Recently, molecular dynamics (MD) and density functional theory (DFT) investigations showed that PhaG membranes present Young’s modulus value $\sim$ 800 GPa (close to the one reported to graphene, $\sim$ 1000 GPa) \cite{pereira2016anisotropic,sun2016mechanical}. Its electronic bandgap can be modified by applying small shear stresses \cite{opengapPHA2016}. Reactive MD simulations revealed that PhaG and graphene membranes have similar fracture behaviors \cite{sui2017MECHprop,CITARNOSSOMRS2018}. Proposed PhaG membranes and nanoribbons applications include Lithium-ion energy storage \cite{Li-ion2018,Li-ion2017} and anti-cancer drug delivery \cite{PHAanticancer2018}. Despite the increasing interest in the physical/chemical properties of PhaG structures, no study about phagraphene nanotubes (PhaNTs) has been reported. 

In the present work, we investigated whether PhaNTs generated from rolling up PhaG membranes is energetically possible. We have used fully-atomistic reactive MD simulations to investigate the structural, dynamical, and mechanical properties of PhaNTs under uniaxial stress at different temperatures. For comparison purposes, we also considered some CNTs of similar diameters.

\section{Computational Methodology}
\label{sec2}

\subsection{On the Generation of Model Phagraphene Nanotubes}
\begin{figure}[pos=ht]
\centering
\includegraphics[width=0.7\linewidth]{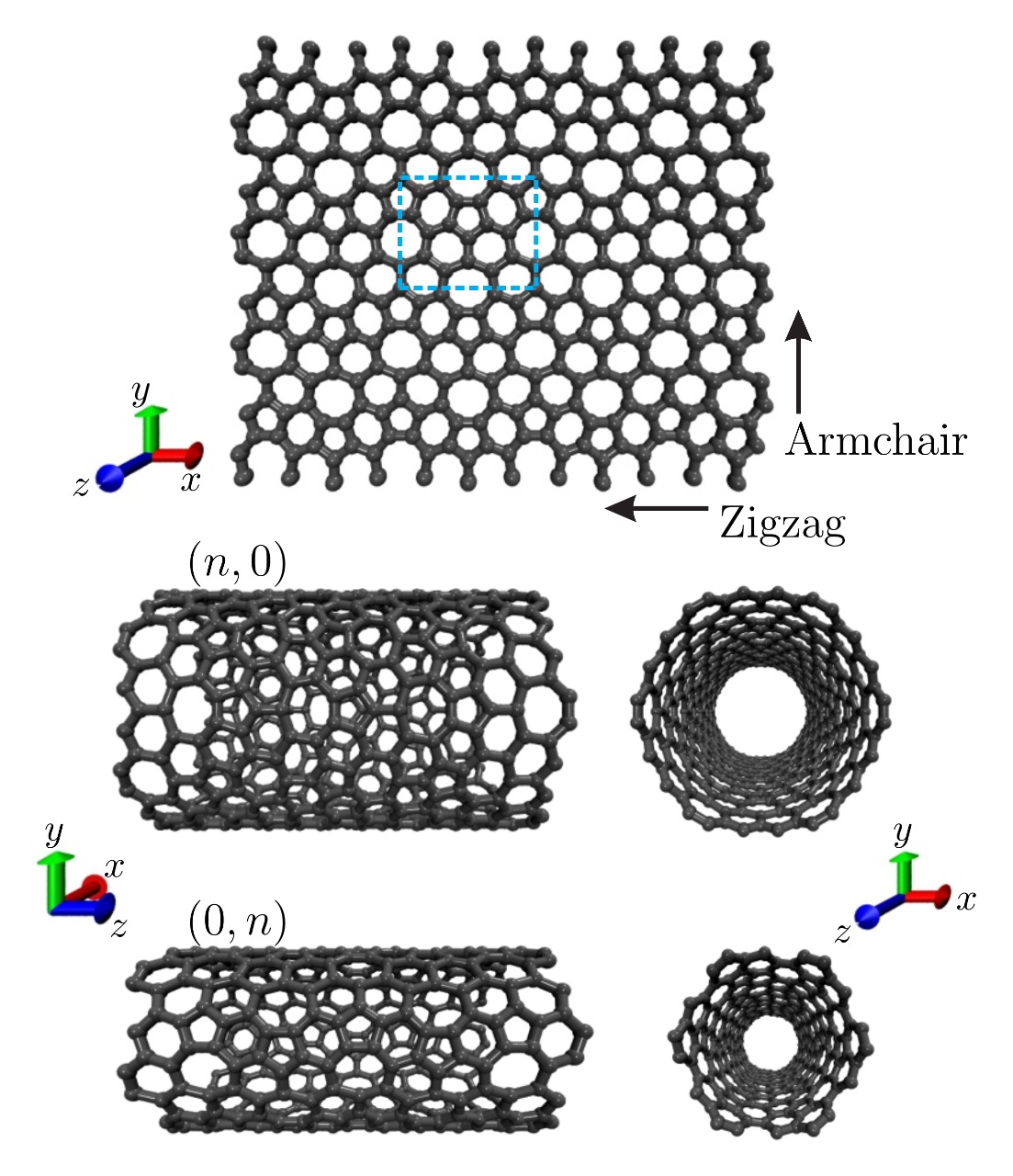} 
\caption{Schematic representation for the PhaG membrane and PhaNTs studied here. Top panel: unit cell of a PhaG membrane; Middle and bottom panels show zigzag and armchair-like PhaNTs, respectively. For the PhaNTs, the reciprocal lattice vectors can be obtained as being $\mathbf{b}_1=b_x(1,0)$ and $\mathbf{b}_2=b_y(0,1)$, with $b_x=2\pi/a_x$ and $b_y=2\pi/a_y$.}
\label{fig1}
\end{figure}

The unit cell of a PhaG membrane ($20$ carbon atoms) can be defined by a rectangle (8.4 \AA~$\times$ 6.4 \AA), as illustrated in Figure \ref{fig1}. In this figure, we also present representative atomistic models of the nanostructures studied here. The top panel shows a PhaG membrane, highlighting its unit cell in a rectangular shape. The middle and bottom panels show zigzag and armchair-like PhaNTs, respectively.

To obtain a PhaNT, we followed the same procedure used to generate standard single-walled CNTs (SWCNTs) \cite{DRESSELHAUS1995883}. The chiral vector, $\mathbf{C}_h$, is defined as   
\begin{eqnarray}
\mathbf{C}_h=(n,m)=n\cdot\mathbf{a}_1+m\cdot\mathbf{a}_2.
\end{eqnarray}
A chiral angle, $\theta_c$, can also be defined as the angle between $\mathbf{C}_h$ and $\mathbf{a}_1$, resulting in
\begin{eqnarray}
\cos\theta_c=\frac{n}{\sqrt{n^2+m^2}}.
\end{eqnarray}
A translational vector can be defined as the smallest vector orthogonal to $\mathbf{C}_h$ in such a way that $\mathbf{T}=(t_1,t_2)=t_1\cdot\mathbf{a}_1+t_2\cdot\mathbf{a}_2$, with $t_1$ and $t_2$ being integers. Using the fact that $\mathbf{C}_h\cdot\mathbf{T}=0$ and choosing $t_2$ to be positive, we obtain $t_1=-m/d\quad\textrm{and}\quad t_2=n/d$, where $d$ is the maximum number such that $t_1$ and $t_2$ are simultaneously integers. Due to the particular ratio $a_x/a_y\approx1.3125$, values can be very high for a given pair $(n,m)$ different from the $(0,m)$ and $(n,0)$ cases. In this way, we limited our calculations only for the $(0,m)$ and $(n,0)$ cases. These last chiral $(0,m)$ and $(n,0)$ cases have translational vectors given by $(1,0)$ and $(0,1)$, respectively. The number of atoms in the nanotube unit cell is given by 20 times the number $\mathcal{N}$ of $a_x\times a_y$ rectangles within the area defined by $\mathbf{C}_h$ and $\mathbf{T}$. Therefore, $\mathcal{N}$ is obtained by dividing $\mathbf{C}_h\times\mathbf{T}$ per $\mathbf{a}_1\times\mathbf{a}_2$, resulting in
\begin{eqnarray}
  \mathcal{N}=\frac{|\mathbf{C}_h\times\mathbf{T}|}{|\mathbf{a}_1\times\mathbf{a}_2|}=
  \frac{(n^2+m^2)}{d}.
\end{eqnarray}
The length $L$ and radius $R$ of the PhaNT as a function of $\mathbf{T}$ and $\mathbf{C}_h$ can be estimated by
\begin{eqnarray}
L=|\mathbf{T}|=\frac{a\sqrt{(n^2+m^2)}}{d}\quad\textrm{and}\quad
R=\frac{|\mathbf{C}_h|}{2\pi}=\frac{a\sqrt{n^2+m^2}}{2\pi},
\end{eqnarray}
respectively. Analogous to the SWCNT case, the PhaNT unit cell is obtained by rolling up the rectangle sector of the PhaG membrane determined by $\mathbf{C}_h$ and $\mathbf{T}$. All the geometrical and structural parameters for the different nanotubes studied here are presented in Table \ref{tab1}. It is worthwhile to mention that here we considered PhaNTs of different diameters and chiralities (see Table \ref{tab1}). CNTs of similar diameters were also considered for comparison purposes. 

\subsection{Reactive Molecular Dynamics Simulations}
The structural, dynamical, and mechanical properties of PhaNTs were studied through MD simulations using the reactive force field ReaxFF potential \cite{van2001reaxff,mueller2010development}, as implemented in the large-scale atomic/molecular massively parallel simulator (LAMMPS) code \cite{plimpton1995fast}. ReaxFF is a force field that allows the formation and breaking of chemical bonds during the dynamics, which is needed to investigate the PhaNT fracture dynamics.

The classical equations of motion were numerically integrated using the velocity-Verlet integrator with a time-step of $0.05$ fs. We increased the tensile stress in the system by changing the length of the nanotube along the periodic direction, considering an engineering strain rate of $10^{-6}$ fs$^{-1}$. The nanostructures were stretched up to their complete fracture, characterized by a permanent and irreversible rupture of the system in two separate pieces. To prevent the existence of initial stresses within the nanostructures, before the stretching procedure they were equilibrated/thermalized within an NVT ensemble at constant temperatures (300 and 900 K) using the Nos\'e-Hoover thermostat \cite{andersen1980molecular,hoover1985canonical} during 100 ps.
The elastic properties, such as Young’s modulus ($Y$), Fracture Strain ($FS$), and Ultimate Strength ($US$) were obtained from the stress-strain curves \cite{de2018mechanical,de2018mechanicalNanotube}. 

The von Mises stress \cite{mises_1913} per atom $k$, is defined as:
\begin{equation}
\centering
\sigma^{k}_{v} = \sqrt{\frac{(\sigma^{k}_{xx} - \sigma^{k}_{yy})^2 + (\sigma^{k}_{yy} - \sigma^{k}_{zz})^2 + (\sigma^{k}_{xx} - \sigma^{k}_{zz})^2 + 6((\sigma^k_{xy})^2+(\sigma^k_{yz})^2+(\sigma^k_{zx})^2)}{2}},
\end{equation}
where $\sigma^k_{xx}$, $\sigma^k_{yy}$, and $\sigma^k_{zz}$ are the components of the normal stress and $\sigma^k_{xy}$, $\sigma^k_{yz}$, and $\sigma^k_{zx}$ are the components of the shear stress. The von Mises stress values provide useful local structural information on the fracture mechanism, once they can determine the region from which the structure has started to yield the fractured lattice. The MD snapshots and trajectories were obtained by using free visualization and analysis software VMD \cite{HUMPHREY199633}.

\begin{table}[pos=ht]
\centering
\caption{Geometrical and structural parameters for the modeled PhaNTs and CNTs.}
\vspace{0.5cm}
\label{tab1}
\begin{tabular}{cccc}
\hline 
PhaNT & Number of Atoms & Radius(\AA) & Length(\AA)\\
\hline
\hline
(2,0) & 160  & 2.67  & 25.60 \\
(4,0) & 320  & 5.35  & 25.60 \\
(6,0) & 480  & 8.02  & 25.60 \\
(8,0) & 640  & 10.70  & 25.60 \\
(10,0) & 800  & 13.37  & 25.60 \\
(12,0) & 960  & 16.04  & 25.60 \\
\hline \\
\hline 
PhaNT & Number of Atoms & Radius(\AA) & Length(\AA)\\
\hline
\hline
(0,2) & 120  & 2.03  & 25.0 \\
(0,4) & 240  & 4.10  & 25.0 \\
(0,6) & 360  & 6.11  & 25.0 \\
(0,8) & 480  & 8.15  & 25.0 \\
(0,10) & 600  & 10.37  & 25.0 \\
(0,12) & 720  & 12.23  & 25.0 \\
\hline \\
\hline
CNT & Number of Atoms & Radius(\AA) & Length(\AA)\\
\hline
\hline
(7,0) & 168  & 2.74  & 25.56 \\
(13,0) & 312  & 5.10  & 25.56 \\
(21,0) & 504  & 8.22  & 25.56 \\
(27,0) & 648  & 10.57  & 25.56 \\
(34,0) & 816  & 13.31  & 25.56 \\
(41,0) & 984  & 16.05  & 25.56 \\
\hline
\end{tabular}
\end{table}

\section{Results and Discussion}
\label{sec3}

\begin{figure}[pos=ht]
\centering
\includegraphics[width=0.4\linewidth]{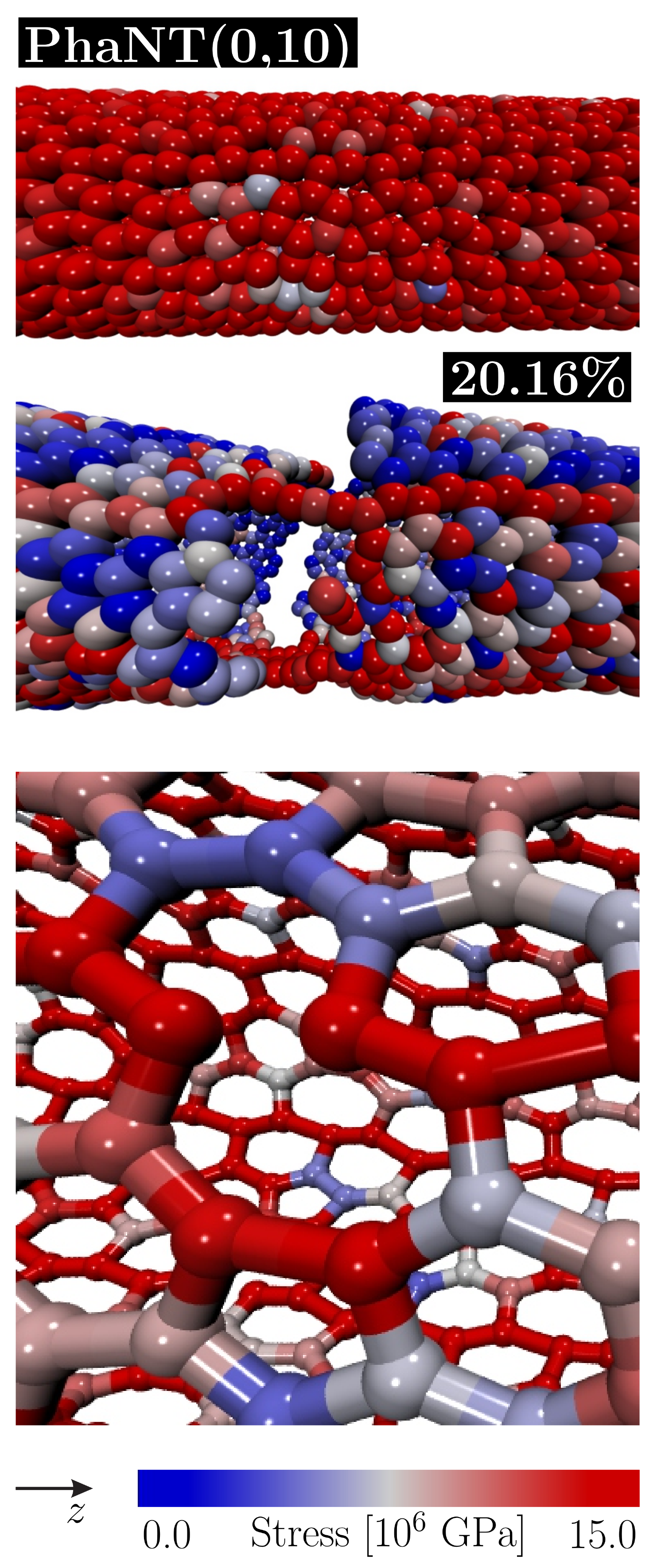}
\includegraphics[width=0.4\linewidth]{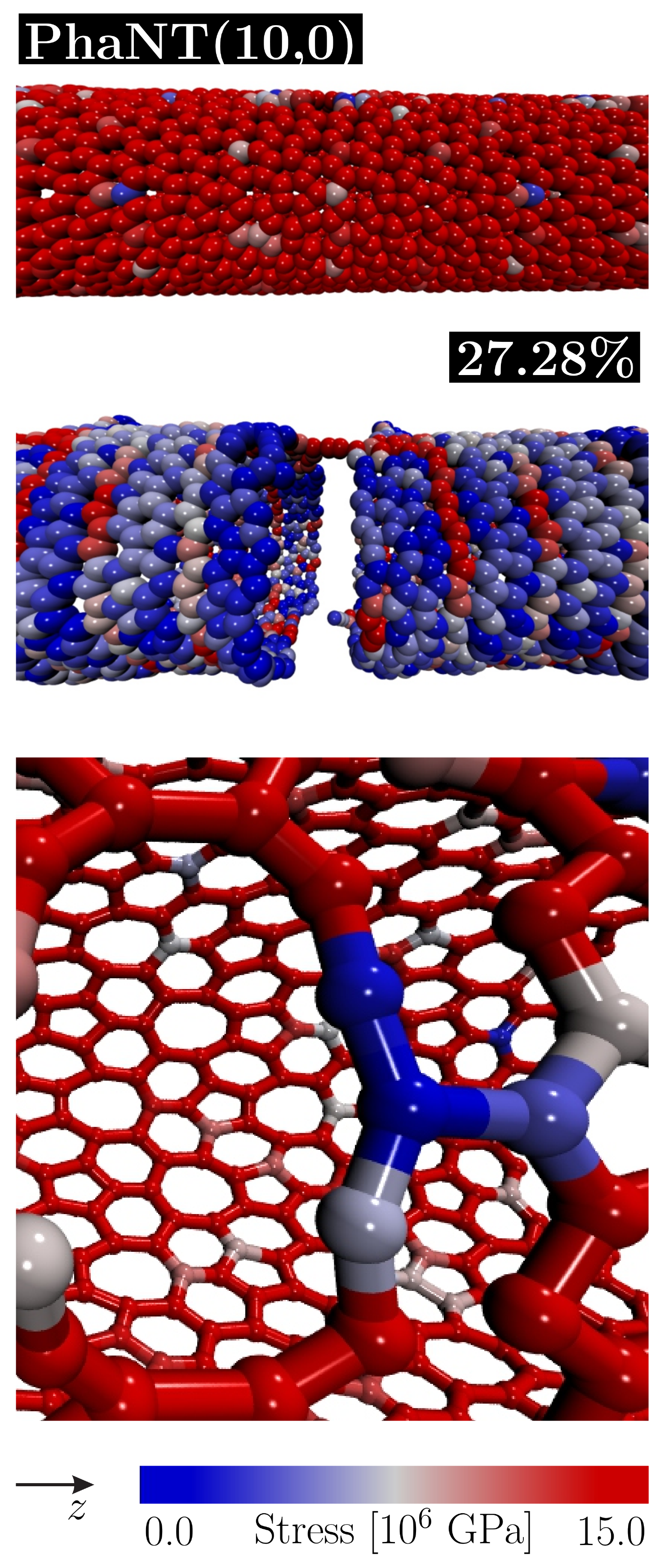}
\caption{Representative MD snapshots for $(0,10)$ (left) and $(10,0)$ PhaNTs (right) at different stress regimes. Results for 300K. The top and middle panels show the stretched structures just before and after structural failure (fracture), respectively. The bottom panels show a zoomed-in view of the regions where the fracture initiates. The structures are colored accordingly to their von Mises stress values, ranging from blue (low stress) to red (high stress).}
\label{fig2}
\end{figure}

In Figure \ref{fig2} we present representative MD snapshots for for $(0,10)$ and $(10,0)$ PhaNTs at different stress regimes. In this figure, we can see the accumulated high and low-stress regions (red and blue colors), with the bonds parallel to the orientation of the applied strain (axial $z$ direction). 

The sequence of panels in Figure \ref{fig2} shows the fracture process of a $(0,10)$ PhaNT (left column) and $(10,0)$ PhaNT (right column) at 300 K. The top panels in Figure \ref{fig2} show MD snapshots for PhaNTs just before fracture, which were found to occur at 20.08 \% and 27.20 of strain for the $(0,10)$ and $(10,0)$ cases, respectively. The middle panels display the complete fractured nanostructures (for 20.16 \% and 27.28 \% of strain for the $(0,10)$ and $(10,0)$ cases, respectively) and the bottom panels zoomed-in a region in which the chemical bond breaking occurs, which is the structural defect that triggers the crack propagation. It is interesting to mention that $(10,0)$ PhaNT presents some wall-buckling even with no applied strain. This buckling is present in all $(n,0)$ PhaNTs studied here but they were not observed for $(0,n)$ PhaNT cases. In contrast to the PhaG membranes \cite{pereira2016anisotropic,de2018mechanical,shirazi2019molecular} for all investigated PhaNT structures we observed $C-C$ bond breaking always parallel to the uniaxial stress direction. Such broken bonds are located on the heptagon-hexagon frontier, as can be seen in the bottom panels of Figure \ref{fig2}. The presence of heptagons makes the PhaNT visibly more porous and also mechanically less stiff when compared to more denser nanostructures, like CNTs \cite{de2019elastic}. The whole process can be better understood from videos 1 and 2 in the Supplementary Materials.

\begin{figure}[pos=ht]
\centering
\includegraphics[width=\linewidth]{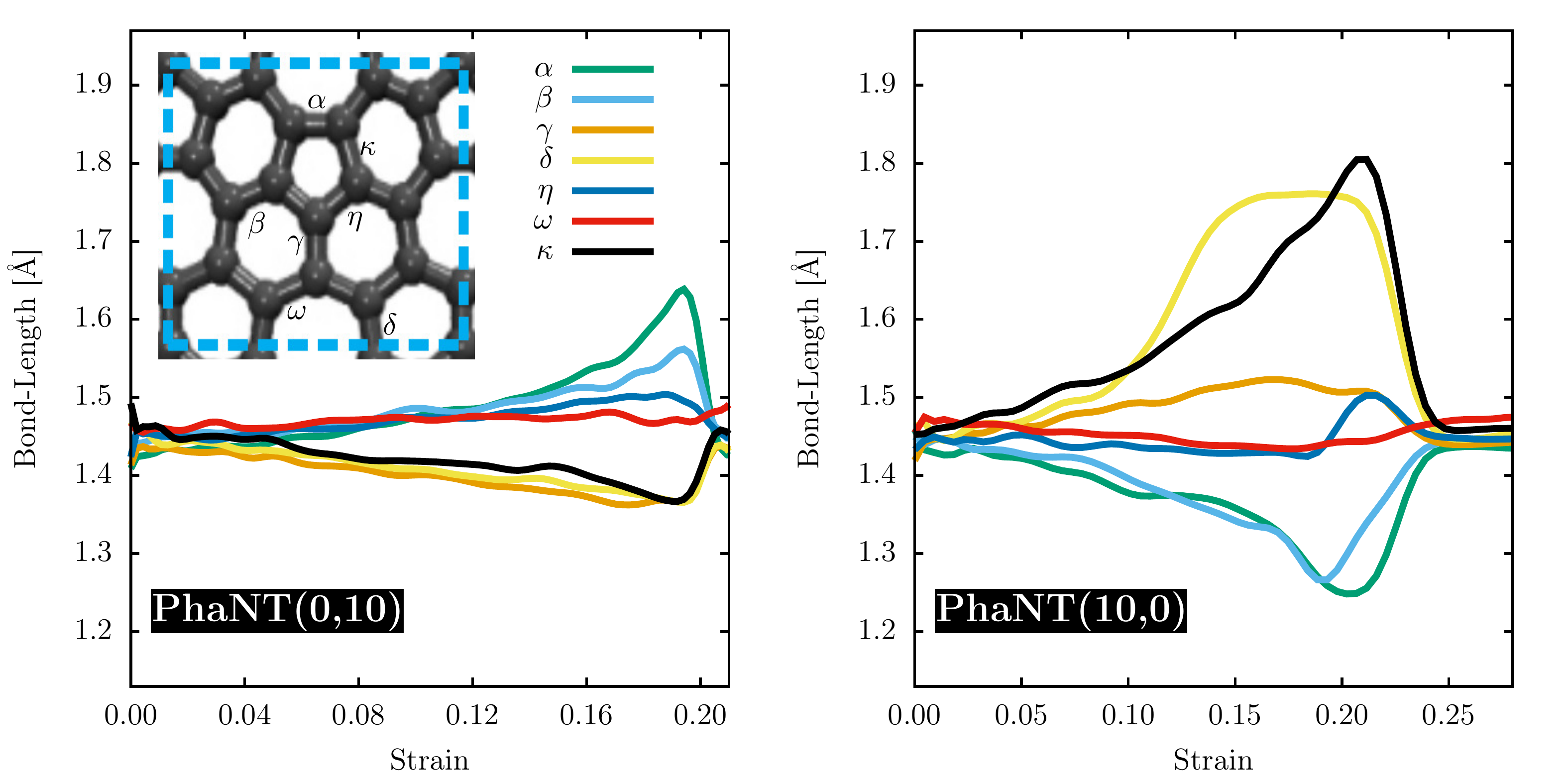} 
\caption{Representative bond-length values as a function of the strain values, for PhaNT (0,10) (left) and PhaNT(10,0) (right) at 300 K.}
\label{fig3}
\end{figure}

In Figure \ref{fig3} we present representative bond-length values as a function of the strain values for (0,10) and (10,0) PhaNTs. The inset shows the bonds, and their respective labels, that are monitored during the stretching process. From this figure, we can see that, except for the $\omega$ bonds, all the other exhibit significant deviations from their initial values. For the PhaNT (0,10) case at 300 K, while the $\alpha$, $\beta$, and $\eta$ bonds are elongated, the $\gamma$, $\delta$, and $\kappa$ ones are compressed. After the tube fracture, all the bonds return to their initial values. For the PhaNT (10,0) case, the behavior is different with large amplitude variations associated with bond breaking ($\sim 1.6$~\AA) and the appearance of carbon triple bonds ($\sim 1.2 $~\AA). The explanation for these distinct behaviors is the same discussed above for Figure \ref{fig2} is are related to the different bonds/rings that are stretched in the (0,n) and (n,0) tube families.

\begin{figure}[pos=ht]
\centering
\includegraphics[width=0.6\linewidth]{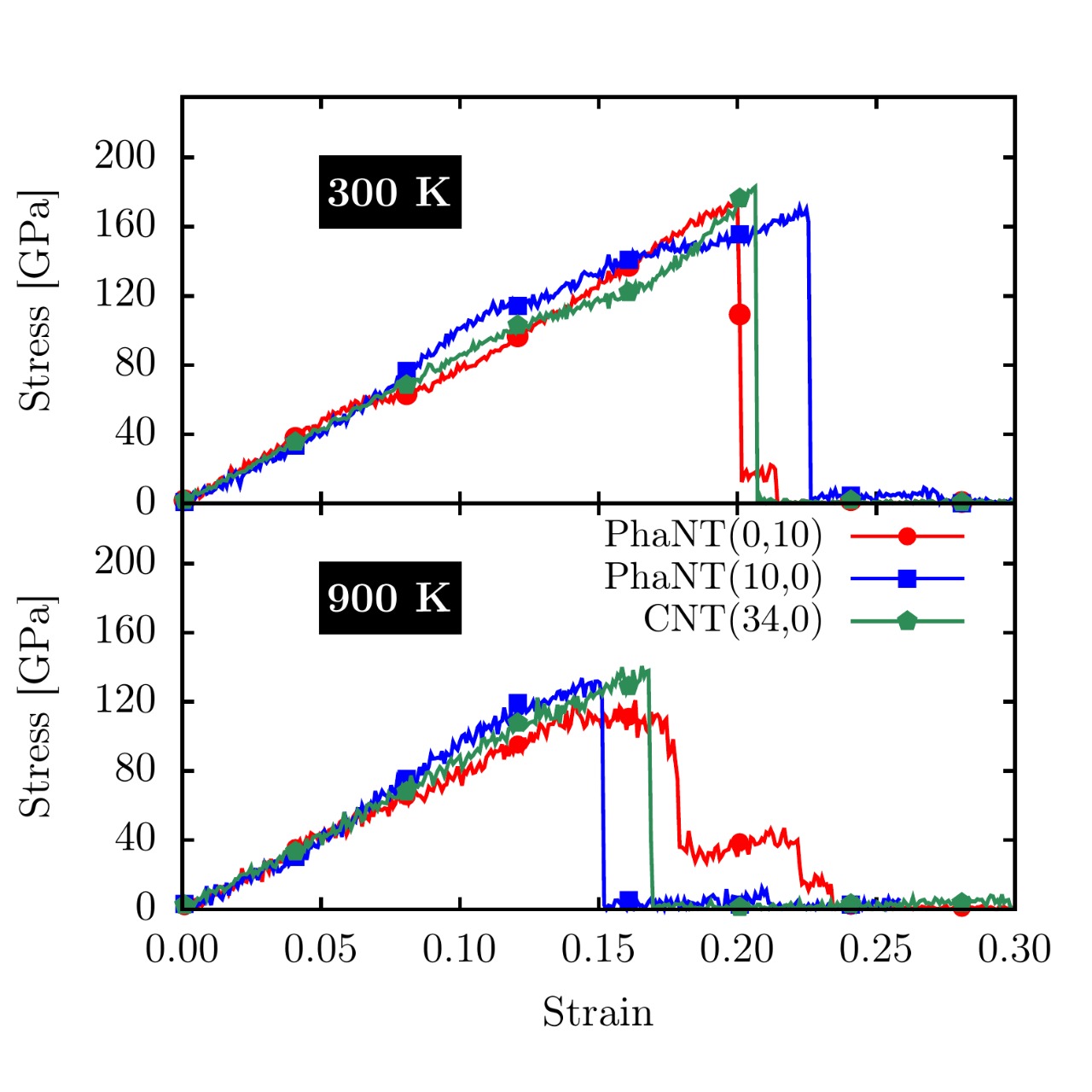} 
\caption{Stress-strain curves for representative cases: ($34,0$) SWCNT, ($10,0$) PhaNT, and ($0,10$) PhaNT at 300 K and 900 K.}
\label{fig4}
\end{figure}

In Figure \ref{fig4}, we present the stress-strain curves for representative CNT (34,0) and PhaNT ((0,10 and (10,0)) cases, which have similar radius values (see Table \ref{tab1}). The results are for 300K (top) and 900K (bottom), respectively. From Figure \ref{fig4} we can see that CNTs and PhaNTs have similar elastic behavior, with almost the same elastic regimes followed by an abrupt drop of the stress values, which can be associated with the fracture occurrence. The obtained average critical strain values are similar at 300K, $\approx 20\%$ for $(0,10)$ PhaNT, $\approx 21\%$ for $(34,0)$ CNT, and $\approx 23\%$ for $(0,10)$ PhaNT. As expected, increasing the temperature to 900K, there is a decrease in the critical tensile strain values for all PhaNTs and CNTs. For $(10,0)$ and $(0,10)$ PhaNTs and $(34,0)$CNT, they significantly decrease from $23\%$ to $15\%$, from $20\% $ to $15\%$, and from $21\%$ to $17\%$, respectively. 

The Young's modulus values were estimated from the linear (lower than $10\%$) regions of the stress-strain curves (see Tables \ref{tab2}-\ref{tab4}). When the temperature is increased to 900 K, no significant changes are observed regarding $Y$ and $FS$. We can conclude that the Young's modulus values of PhaNTs and CNTs are in the same range and they are not very sensitive to thermal effects in the temperature range considered here. As expected, there is a dependence of the structural failure with the temperature, in agreement with the literature \cite{wei2003tensile}. This behavior can be explained by the fact that high temperatures increase atomic amplitude vibrations, which favors bond breaking. As expected, the $US$ also decrease from low to high temperatures. Interestingly, the ordering of the fracture strain values (among different PhaNT chiralities) can be reversed going from low to high temperatures for some tubes. 

\begin{table}[]
\begin{tabular}{|c|c|c|c|c|}
\hline
\multicolumn{5}{|c|}{PhaNT ($n$, 0)}                                           \\ \hline
Temperature {[}K{]}  & ($n$, 0) & $Y_M$ {[}GPa{]} & FS {[}\%{]} & US {[}GPa{]} \\ \hline
\multirow{6}{*}{300} & (2, 0)   & 706.15          & 17.76       & 170.33       \\ \cline{2-5} 
                     & (4, 0)   & 734.57          & 20.24       & 164.61       \\ \cline{2-5} 
                     & (6, 0)   & 809.85          & 17.36       & 158.33       \\ \cline{2-5} 
                     & (8, 0)   & 792.80          & 22.87       & 145.00       \\ \cline{2-5} 
                     & (10, 0)  & 824.59          & 22.24       & 171.11       \\ \cline{2-5} 
                     & (12, 0)  & 796.75          & 22.24       & 169.58       \\ \hline
\multirow{6}{*}{900} & (2, 0)   & 693.47          & 13.52       & 167.71       \\ \cline{2-5} 
                     & (4, 0)   & 726.01          & 15.36       & 149.92       \\ \cline{2-5} 
                     & (6, 0)   & 776.33          & 15.20       & 144.50       \\ \cline{2-5} 
                     & (8, 0)   & 782.01          & 15.89       & 124.00       \\ \cline{2-5} 
                     & (10, 0)  & 763.27          & 14.40       & 133.74       \\ \cline{2-5} 
                     & (12, 0)  & 764.19          & 14.16       & 135.01       \\ \hline
\end{tabular}
\label{tab2}
\caption{Elastic properties for ($n$, 0) CNTs: Young's modulus ($Y_M$), fracture strain (FS), and ultimate strength (US).}
\end{table}

\begin{table}[]
\begin{tabular}{|c|c|c|c|c|}
\hline
\multicolumn{5}{|c|}{PhaNT (0, $n$)}                                           \\ \hline
Temperature {[}K{]}  & (0, $n$) & $Y_M$ {[}GPa{]} & FS {[}\%{]} & US {[}GPa{]} \\ \hline
\multirow{6}{*}{300} & (0, 2)   & 826.64          & 17.92       & 168.61       \\ \cline{2-5} 
                     & (0, 4)   & 978.92          & 18.96       & 176.71       \\ \cline{2-5} 
                     & (0, 6)   & 965.87          & 20.08       & 176.53       \\ \cline{2-5} 
                     & (0, 10)  & 937.21          & 19.92       & 174.43       \\ \cline{2-5} 
                     & (0, 12)  & 945.51          & 19.52       & 176.03       \\ \hline
\multirow{6}{*}{900} & (0, 2)   & 805.54          & 12.96       & 126.07       \\ \cline{2-5} 
                     & (0, 4)   & 923.82          & 16.40       & 124.60       \\ \cline{2-5} 
                     & (0, 6)   & 899.93          & 13.20       & 115.69       \\ \cline{2-5} 
                     & (0, 10)  & 914.64          & 16.32       & 120.90       \\ \cline{2-5} 
                     & (0, 12)  & 899.75          & 15.12       & 114.36       \\ \hline
\end{tabular}
\label{tab3}
\caption{Elastic properties for (0, $n$) CNTs: Young's modulus ($Y_M$), fracture strain (FS), and ultimate strength (US).}
\end{table}

\begin{table}[]
\begin{tabular}{|c|c|c|c|c|}
\hline
\multicolumn{5}{|c|}{CNT ($n$, 0)}                                             \\ \hline
Temperature {[}K{]}  & ($n$, 0) & $Y_M$ {[}GPa{]} & FS {[}\%{]} & US {[}GPa{]} \\ \hline
\multirow{6}{*}{300} & (7, 0)   & 972.69          & 14.72       & 244.27       \\ \cline{2-5} 
                     & (13, 0)  & 931.83          & 20.24       & 244.14       \\ \cline{2-5} 
                     & (21, 0)  & 923.47          & 20.64       & 240.87       \\ \cline{2-5} 
                     & (27, 0)  & 924.51          & 20.32       & 226.36       \\ \cline{2-5} 
                     & (34, 0)  & 970.39          & 20.40       & 226.05       \\ \cline{2-5} 
                     & (41, 0)  & 916.75          & 20.16       & 219.76       \\ \hline
\multirow{6}{*}{900} & (7, 0)   & 965.59          & 12.16       & 188.44       \\ \cline{2-5} 
                     & (13, 0)  & 921.32          & 14.80       & 181.89       \\ \cline{2-5} 
                     & (21, 0)  & 892.42          & 15.36       & 162.43       \\ \cline{2-5} 
                     & (27, 0)  & 889.70          & 15.84       & 171.70       \\ \cline{2-5} 
                     & (34, 0)  & 926.57          & 16.32       & 170.41       \\ \cline{2-5} 
                     & (41, 0)  & 877.96          & 16.80       & 174.36       \\ \hline
\end{tabular}
\label{tab4}
\end{table}

\section{Conclusions}
\label{sec4}

In summary, we have carried out fully-atomistic reactive (ReaxFF) simulations to study the mechanical properties of phagraphene tubes (PhaNTs). For comparison purposes, the results were also contrasted to similar (in terms of diameters and chiralities) carbon nanotubes (CNTs). An axial strain (with periodic boundary conditions along the $z-direction$) was applied to the nanotubes at two different temperatures: 300 and 900 K. Our results show that for all PhaNTs, the fracture starts in the largest rings formed by C-C bonds connecting heptagons and hexagons. In contrast to the PhaG membranes, for the PhaNTs the $C-C$ bond breaking is always parallel to the uniaxial stress direction. The Young's modulus values obtained for PhaNTs were smaller (but in the same magnitude) than the corresponding ones for CNTs. The values for 300K and 900K are only slightly different. These differences can be explained by the different atomic configurations that are present in PhaNTs (which is more porous due to the presence of $5-6-7$ rings). Both, PhaNTs and CNTs, present equivalent fracture strains ranging between 15\%-20\%. For the ultimate strength values, CNTs present values about 30\% higher than the corresponding ones for PhaNTs.       

\section*{Acknowledgements}
This work was supported by CAPES, CNPq, FAPESP, FAPDF, and ERC and the Graphene FET Flagship. J.M.S., R.A.B., and D.S.G. thank the Center for Computational Engineering and Sciences at Unicamp for financial support through the FAPESP/CEPID Grants $2013/08293-7$ and \#2018/11352-7. L.A.R.J., A.L.A, and W.H.S.B. acknowledges CENAPAD-SP for computer time. A. L. A. acknowledges CNPq (Process No. 427175/20160) for financial support. W.H.S.B. and A.L.A. thank the Laborat\'orio de Simula\c c\~ao Computacional Caju\'ina (LSCC) at Universidade Federal do Piau\'i for computational support. Finally, L.A.R.J acknowledges the financial support from a Brazilian Research Council FAPDF and CNPq grants $00193.0000248/2019-32$ and $302236/2018-0$, respectively. 

\printcredits
\bibliographystyle{unsrt}
\bibliography{cas-refs}

\end{document}